\documentstyle[12pt,epsf]{article}
\newcommand{\be}{\begin{equation}}
\newcommand{\ee}{\end{equation}}
\newcommand{\bea}{\begin{eqnarray}}
\newcommand{\eea}{\end{eqnarray}}
\newcommand{\bt}{\begin{tabular}}
\newcommand{\et}{\end{tabular}}
\newcommand{\ba}{\begin{array}}
\newcommand{\ea}{\end{array}}
\newcommand{\ov}{\overline}

\newcommand{\bvec}{\mathbf}
\newcommand{\gr}{ \nabla \, }
\def\ne{\hbox{$\nu_e \!$ }}
\def\nm{\hbox{$\nu_\mu \!$ }}
\def\nt{\hbox{$\nu_\tau \!$ }}

\def\rt{\hbox{$\rightarrow$ }}

\newcommand{\vem}{V_{eff}^{e^-}}
\newcommand{\vep}{V_{eff}^{e^+}}
\newcommand{\vp}{ V_{eff}^{p}}
\newcommand{\vn}{ V_{eff}^{n}}
\newcommand{\fem}{{\bvec{{\cal F}}}_{\nu}^{e^-}}
\newcommand{\fep}{{\bvec{{\cal F}}}_{\nu}^{e^+}}
\newcommand{\fp}{ {\bvec{{\cal F}}}_{\nu}^{p} }
\newcommand{\fn}{ {\bvec{{\cal F}}}_{\nu}^{n} }

\begin{document}
\setlength{\unitlength}{1mm}

\hfill{
\begin{tabular}{l}
DSF$-$98/39 \\ INFN$-$NA$-$IV$-$98/39 \\
hep-ph/9902357
\end{tabular}}
\vspace*{2cm}

\begin{center}
\begin{Large}
{\bf The action of neutrino ponderomotive force on Supernova dynamics }
\end{Large}
\end{center}

\bigskip\bigskip

\begin{center}
{\Large
Salvatore Esposito
\footnote{e-mail: sesposito@na.infn.it} }
\end{center}

\vspace{.5cm}

\begin{center}
{\it
\noindent
Dipartimento di Scienze Fisiche, Universit\`{a} di Napoli ``Federico II''\\ and
\\ Istituto Nazionale di Fisica Nucleare, Sezione di Napoli\\ Mostra
d'Oltremare Pad. 20, I-80125 Napoli, Italy }
\end{center}

\bigskip\bigskip\bigskip
\begin{abstract}
\noindent
Collective interactions of a beam of neutrinos/antineutrinos traversing a
dense magnetized plasma of electrons/positrons, protons and neutrons are
studied with particular reference to the case of a Supernova. We find that
the ponderomotive force exerted by neutrinos gives, contrary to
expectations, a negligible contribution to the revival of the shock for a
successful Supernova explosion, although new types of convection and plasma
cooling processes induced by the ponderomotive force could be, in
principle, relevant for the dynamics itself.
\end{abstract}

\vspace*{2cm}


\baselineskip=.8cm

\newpage
\noindent
At the end of their lives, massive stars ($M > 8 M_\odot$) develop into
Supernovae of type II and, after explosion, neutron stars are born. The
mechanism of type II Supernovae is approximately well understood, although
some obscure points remain \cite{Bethe}, \cite{Burrows}.

The iron core of the progenitor star has a mass around the Chandrasekhar
limit ($\sim 1.4 M_\odot$) so it has no stable configuration and will
collapse. This is also accelerated by the negative nuclear pressure. During
the collapse, the key process is the electron capture by free protons and
transitions in complex nuclei, producing (electron) neutrinos that escape
from  the star, thus deleptonizing it. Electron capture would continue
indefinitely, and the final electron fraction $Y_e^f$ at collapse would be
extremely small, were it not for the trapping of neutrinos taking place
when the density in the core has reached very high values ($\rho \sim
10^{11} \div 10^{12} \, g/cm^3$). After neutrinos are trapped, due to (weak
neutral current mediated) elastic scattering by nuclei, they fill available
phase space and their distribution can be approximately described by a
Fermi-Dirac distribution with effective temperature $T_\nu$ and chemical
potential $\mu_\nu$ (different from those of electrons). The collapse can
be stopped only when the nuclear pressure becomes positive, which happens
when the nuclei touch and fuse together, forming nuclear matter. Thus,
collapse continues until the central density becomes substantially greater
than nuclear density ($\sim (3 \div 4) {\times} 10^{15} \, g/cm^3$). At this
point, the inner core rebounds and sends a shock wave out into the star.
The shock forms not in the centre of the star, but near the surface of the
homologous core (slightly outside it). \\ In few cases, if both the
temperature at the beginning of the collapse and the mass of the iron core
are low \cite{Bruenn}, the shock will go through the entire star and expel
most of it, thus giving a Supernova ({\it prompt mechanism}). The core
remains behind and will become a neutron star; the large negative
gravitational energy of that neutron star provides the energy to expel the
mantle and the envelope against gravitational attraction of the core. The
major part of the released gravitational energy of the neutron star goes
into the emission of neutrinos, which mainly occurs after the Supernova
material has been set in outward motion. \\ Behind the shock, nuclei are
dissociated into nucleons at a cost in energy of about $8.8 \, MeV$ per
nucleon, and this gradually drains the energy of the shock. If the amount
of material the shock has to traverse from the homologous core before it
emerges from the iron core is not small, the prompt mechanism fails,
because the shock rapidly looses energy to dissociate nuclei into nucleons.
The shock is therefore apt to stall at some point (typically at $r \sim 400
\, Km$) and then it turns into an accretion shock in which additional
infalling matter accretes to the existing core. The outward motion has then
stopped and the prompt shock has failed to expel the outer part of the
star. Furthermore, once the shock slows down, there is time for copious
neutrino emission, which further saps its energy. However, if the shock
turns into an accretion shock, neutrinos from the core can be absorbed
(after about $0.5 \div 1 \, s$) by material at $r \sim 100 \div 200 \, Km$,
and can heat this sufficiently to revive the shock, which will then expel
the material from the star ({\it delayed mechanism}) \cite{Wilson}.\\
However, in most cases, the outgoing material (ejecta) is too far away to
effectively absorb the neutrinos from the core but, in numerical
simulations, it is found that between ejecta and the core there is a low
density region ($\rho \sim 10^7 \, g/cm^3$) filled mainly by radiation (the
radiation bubble) \cite{Cooperstein}. In this bubble, $e^+$ $e^-$ pairs can
absorb neutrino energy by elastic scattering (also $\nu \ov{\nu} \rt e^+
e^-$ can occur), thus it continually receives new energy and continually
exerts pressure on the ejecta, then driving the shock. We have to note that
density increases outward near the bubble outer surface, so that dilute
material in the bubble has to support and push down material in the ejecta.
This causes Rayleigh-Taylor instability and convection will occur. In the
delayed mechanism, convection is very fast (it is practically instantaneous
compared to diffusion time scales) and reduces the energy loss by
re-emission of neutrinos. Furthermore, there is an important effect. The
absorption of neutrino energy takes place mainly at $r \sim 100
\div 200 \, Km$; the convection brings this energy to the shock front,
whether that is at $r \sim 300 \, Km$ or $r \sim 4000 \, Km$, and thereby
continually supports the shock. This will succeed in ejecting the material
outside the bubble.

These are the essential features of Supernova dynamics, although there are
still some doubts especially on how the explosion mechanism can undergo
effectively. In particular, it is clear that neutrinos are responsible for
transporting energy from the central core to the layers surrounding it, but
it seems that the collisional loss rate is marginal to produce the required
heating. Moreover, besides the key role played by neutrinos, a complete
understanding of hydrodynamic instabilities and overturn before, during and
after core collapse is also crucial.

In this paper we will concentrate  on the possible collective interactions
between neutrinos and the dense plasma in connection with the problem of
the heating of the shocked envelope in a Supernova, as proposed in
\cite{Bingham},\cite{b2}. The idea is that the material in the shock is
continually run over by an intense neutrino flux, so that stimulated
scattering processes can take place, in analogy with stimulated Raman or
Compton scattering for laser coupling to plasma oscillations. Thus we first
study the (weak interaction) ponderomotive force exerted by neutrinos on
the background electrons, positrons, protons and neutrons, and then apply
the results to the case of Supernova explosion.

\vspace{1cm}

Let us consider neutrinos with momentum $\bvec{p}_\nu$ whose phase space
distribution is $f_\nu(t,\bvec{r},\bvec{p}_\nu)$ ($(t,\bvec{r})$ is the
4-position), acting on the background particles $\alpha \, = \, e^{\pm} , p ,n$
of a plasma with momentum $\bvec{p}_\alpha$, distributed according to
$f_\alpha(t,\bvec{r},\bvec{p}_\alpha)$, through an effective potential
\begin{equation}\label{1}
  {\cal V}_{eff}^\alpha \; = \; 2 \, \int \, \frac{d \bvec{p}_\alpha}
  {(2 \pi)^3} \,\, V_{eff}^\alpha (t,\bvec{r},\bvec{p}_\alpha)
\end{equation}
(the factor 2 is the statistical weight for the fermions $\alpha \, = \,
e^{\pm} , p ,n$). The force exerted by neutrinos over (a single) background
particle $\alpha
\, = \, e^{\pm} , p ,n$ can then be written in the form \cite{pond}
\bea
\bvec{{\cal F}}_\nu^\alpha & = & \frac{1}{n_\alpha} \, 2 \,
\int \, \frac{d \bvec{p}_\nu}{(2 \pi)^3} \, \int \, \frac{d
\bvec{p}_\alpha} {(2 \pi)^3} \, \left( f_\nu(t,{\bvec{r}},{\bvec{p}}_\nu)
 \, \gr V_{eff}^\alpha (t,\bvec{r},\bvec{p}_\alpha) \; + \right.
 \nonumber \\
 & - & \left. \gr \left( f_\nu(t,{\bvec{r}},{\bvec{p}}_\nu) \, \left(
 \frac{\partial V_{eff}^\alpha}{\partial f_\alpha} \right)_T \,
 f_\alpha(t,{\bvec{r}},{\bvec{p}}_\alpha) \right) \, \right)
\label{2}
\eea
where the particle number densities are
\bea
n_\alpha(t, \bvec{r}) & = & 2 \, \int \, \frac{d \bvec{p}_\alpha}{(2
\pi)^3} \, f_\alpha(t,\bvec{r},\bvec{p}_\alpha) \label{3} \\
n_\nu(t, \bvec{r}) & = & \int \, \frac{d \bvec{p}_\nu}{(2
\pi)^3} \, f_\nu(t,\bvec{r},\bvec{p}_\nu) \label{4}
\eea
(neutrinos are completely polarized particles). We now have to calculate
the effective potential for $\nu - \alpha$ coherent interactions. Let us
first note that in Supernovae large magnetic fields $\bvec{B}$ are, in
general, present \cite{BB} so that neutrino interaction with the background
is modified with respect to the case in which the medium is non magnetized.
However, since the calculation of the effective potential proceeds from
that of neutrino self-energy in the considered medium \cite{NR},
\cite{magn}, \cite{magn2}, in general one can write
\begin{equation}\label{5}
  V_{eff}^{tot} \; = \; V_{eff}^{\bvec{B} = 0} \; + \; V_{eff}^{\bvec{B}
  \neq 0}
\end{equation}
where the first term refers to the $\bvec{B} = 0$ case while the second one
explicitly involves the presence of the magnetic field. \\ Let us first
examine the contribution independent on $\bvec{B}$ \footnote{In what
follows we consider only isotropic media or, more in general, anisotropic
ones with equal fluxes of electrons moving in opposite directions (see
\cite{magn2}).}. Two kind of Feynman diagrams contribute to neutrino
self-energy \cite{NR}: one involves charged current interaction between \ne
and electrons while in the others we are concerned with neutral current
interactions between each neutrino flavour and electrons, protons and
neutrons. For neutrino-electron interactions we have
\begin{equation}\label{6}
  \vem \; = \; \sqrt{2} \, G_F \, g_V \, f_{e^-}(t,\bvec{r},\bvec{p}_{e^-})
\end{equation}
where $G_F$ is the Fermi coupling constant and
\bea
g_V & = & g_V^{CC} \; + \; g_V^{NC} \;\;\;\;\;\;\;\;\;\;\;\;\;\;\; {\mathrm
for} \; \ne \label{6a} \\ g_V & = & g_V^{NC} \;\;\;\;\;\;\;\;\;\;\;\;\,
\;\;\;\;\;\;\;\;\;\;\;\;\;\;\; {\mathrm for} \; \nm \, , \, \nt
\label{6b}
\eea
\be
g_V^{CC} \; = \; 1 \;\;\;\;\;\;\;\;\;\; , \;\;\;\;\;\;\;\;\;\; g_V^{NC}
\; = \; 2 \, \sin^2
\,
\theta_W \; - \; \frac{1}{2} \label{6c}
\ee
with $\theta_W$ the Weinberg angle of the electro-weak standard model. The
effective potentials for neutrino-positron and neutrino-proton are instead
given by
\begin{eqnarray}
  \vep & = & - \, \sqrt{2} \, G_F \, g_V \, f_{e^+}
  (t,\bvec{r},\bvec{p}_{e^+}) \\
  \vp & = & - \, \sqrt{2} \, G_F \, g_V^{NC} \, f_p
  (t,\bvec{r},\bvec{p}_p) ~~~~.
\end{eqnarray}
Finally, neutrino-neutron interactions are described by
\begin{equation}\label{8}
  \vn \; = \; - \, \frac{G_F}{\sqrt{2}} \, g_V  \, f_n
  (t,\bvec{r},\bvec{p}_n) ~~~~.
\end{equation}
Note also that the effective potential for antineutrino interactions is
opposite to that for neutrino. Then, from Eq.(\ref{2}), we obtain the
$\bvec{B} = 0$ contribution to the ponderomotive force experienced by
$e^{\pm}$, $p$, $n$ due to a beam of neutrinos/antineutrinos:
\bea
\fem & = & - \, \fep \; = \; - \, \sqrt{2} \, G_F \, \left(
\gr (n_{\nu_e} \, - \, n_{\ov{\nu}_e}) \; + \; g_V^{NC} \,
\gr (n_{\nu} \, - \, n_{\ov{\nu}}) \right) \label{9} \\
\fp & = & \sqrt{2} \, G_F \, g_V^{NC} \,
\gr (n_{\nu} \, - \, n_{\ov{\nu}}) \label{10} \\
\fn & = & \frac{G_F}{\sqrt{2}} \, \gr (n_{\nu} \, - \, n_{\ov{\nu}})
\label{11}
\eea
($n_{\nu} \, = \, n_{\nu_e} \, + \, n_{\nm} \, + \, n_{\nt}$). \\ Let us
now consider the contribution induced by the presence of the magnetic field
(the second term in (\ref{5})), which we suppose to lie in the positive $z$
direction. For neutrino-electron interactions, by indicating with
$\lambda_z$ the polarization of the electrons in the plasma, we have
\cite{magn2}
\begin{equation}\label{12}
  \vem \; = \; - \, \sqrt{2} \, G_F \, g_A \, \cos \, \alpha_B \,
  \lambda_z \, f_e(p_{ez},n,\lambda_z)
\end{equation}
where $\cos \, \alpha_B \, = \, \hat{\bvec{p}}_\nu {\cdot} \hat{\bvec{B}}$ and
\bea
g_A & = & g_A^{CC} \; + \; g_A^{NC} \;\;\;\;\;\;\;\;\;\;\;\;\;\;\; {\mathrm
for} \; \ne  \\ g_A & = & g_A^{NC} \;\;\;\;\;\;\;\;\;\;\;\;\,
\;\;\;\;\;\;\;\;\;\;\;\;\;\;\; {\mathrm for} \; \nm \, , \, \nt
\eea
\begin{equation}
g_A^{CC} \; = \; - \, 1 \;\;\;\;\;\;\;\;\;\;\;\; g_A^{NC} \; = \;
\frac{1}{2} ~~~~.
\end{equation}
In (\ref{12}), $f_e(p_{ez},n,\lambda_z)$ is the electron distribution
function (for simplicity we have suppressed the spatial dependence) which
for a magnetized medium takes the form
\begin{equation}\label{13}
  f_e(p_{ez},n,\lambda_z) \; = \; \left( 1 \; + \; \exp \left\{
  \frac{\epsilon(p_{ez},n,\lambda_z) \, - \, \mu}{T} \right\} \right)^{-1}
\end{equation}
where $\epsilon(p_z,n,\lambda_z)$ are the quantized Landau energy levels
given by ($e>0$)
\begin{equation}\label{14}
  \epsilon(p_{ez},n,\lambda_z) \; = \; \sqrt{p_{ez}^2 \; + \; m_e^2 \;
  + e \, B \, \left( 2n \, + \, 1 \, \, + \,  \lambda_z \right) }
\end{equation}
with $n = 0,1,2,...$, $\lambda_z = {\pm} 1$. For neutrino-positron interactions
the same form in (\ref{12}) holds, but with the replacements $V_{eff} \rt
- V_{eff}$ and $\mu \rt - \mu$.
Instead for neutrino-nucleon interactions the effective potential is still
proportional to the polarization of the considered background particle, but
now, for the case of interest in a Supernova, in which nucleons are
strongly (or not weakly) non degenerate, this will be given by the proton
$\mu_p \, = \, 2.79 \, e/2 M_N$ or neutron $\mu_n \, = \, - 1.91\,e/2 M_N$
($M_N$ is the nucleon mass) magnetic moments. We therefore have
\cite{magn2}
\bea
{\cal V}_{eff}^p & = & - \, \frac{G_F}{\sqrt{2}} \, \frac{C_A^N}{C_V^N}
\, n_p \, \frac{\mu_p B}{T} \, \cos \, \alpha_B   \label{15} \\
{\cal V}_{eff}^n & = &  \frac{G_F}{\sqrt{2}} \, \frac{C_A^N}{C_V^N}
\, n_n \,  \frac{\mu_n B}{T} \, \cos \, \alpha_B \label{16}
\eea
where $C_A^N/C_V^N \, \simeq \, 1.26$ is the axial to vector nucleon form
factor. \\ Let us now pass to calculate the ponderomotive force induced by
the effective potential in (\ref{12}); first observe that, since the energy
levels are quantized, we have
\begin{equation}\label{17}
  2 \, \int \, \frac{d \bvec{p}_e}{(2 \pi)^3} \;\;\; \longrightarrow \;\;\;
  \sum_{n = 0}^\infty \, \sum_{\lambda_z} \, 2 \pi \, e B \,
  \int_{- \infty}^\infty \, \frac{d p_{ez}}{(2 \pi)^3} ~~~~.
\end{equation}
Then from Eq. (\ref{2}) we get
\begin{equation}\label{18}
  \fem \; = \; \sqrt{2} \, G_F \, g_A \, \sum_{n = 0}^\infty \,
  \sum_{\lambda_z} \, 2 \pi \, \frac{e B}{(2 \pi)^2} \;
  \int_{- \infty}^\infty \, d p_{ez} \; \lambda_z \, f_e(p_{ez},n,
  \lambda_z) \; \int \, \frac{d \bvec{p}_\nu}{(2 \pi)^3} \; \cos \,
  \alpha_B \, \gr f_\nu ~~~~.
\end{equation}
Since $f_\nu$ does not depend on neutrino angles,
\begin{equation}\label{19}
  \int \, \frac{d \bvec{p}_\nu}{(2 \pi)^3} \; \cos \, \alpha_B \,
  \gr f_\nu \; = \; \frac{1}{(2 \pi)^3} \, \int \, p_\nu^2 \, d p_\nu
  \, \gr f_\nu \, \int_{- 1}^1 \, d \cos \, \alpha_B \, \cos \, \alpha_B
  \; = \; 0
\end{equation}
and so
\begin{equation}\label{20}
  \fem \; = \; 0 ~~~~.
\end{equation}
In a similar fashion, we can also prove that
\begin{equation}\label{21}
  \fep \; = \; \fp \; = \; \fn \; = \; 0 ~~~~.
\end{equation}
We then have the important result that the magnetic field dependent term in
the effective potential (\ref{5}) does not contribute to the ponderomotive
force of neutrinos by virtue of the dependence of $V_{eff}^{\bvec{B}} \neq
0$ on $\cos \, \alpha_B \, = \, \hat{\bvec{p}}_\nu {\cdot} \hat{\bvec{B}}$, and
thus Eqs. (\ref{9})-(\ref{11}) are general results. \\ Some considerations
are now in order. Due to weak couplings, we have that, as regards nucleons,
\begin{equation}\label{22}
  | \fp | \; \approx \; 8 \% \; | \fn |
\end{equation}
so the neutrino force on neutrons is more efficient than that on protons;
secondly, for a given neutrino net number, the neutral current contribution
to $\bvec{{\cal F}}_\nu^e$ is about 4 \% of that of charged current one. \\
Considering the typical situation present in a Supernova, we can also
predict the sign of the ponderomotive force exerted by neutrinos. For
$\nm$, $\nt$ the neutrino-nucleon elastic scattering cross section is
(slightly) higher than the one for antineutrino-nucleon \cite{charlie}, so
antineutrinos escape more easily from the star, leaving the Supernova more
rich in neutrinos:
\begin{equation}\label{23}
  n_{\nu_{\mu , \tau}} \; - \; n_{\ov{\nu}_{\mu , \tau}} \; > \; 0 ~~~~.
\end{equation}
The same approximately holds for the thermally produced $\ne, \ov{\nu}_e$;
moreover there are also $\ne$ from the deleptonization of the Supernova
core, and then
\begin{equation}\label{24}
  n_{\nu_{e}} \; - \; n_{\ov{\nu}_{e}} \; > \; 0 ~~~~.
\end{equation}
Then, the net neutrino number density is a decreasing function of the
distance $r$ from the centre of the star, so that its gradient is negative:
\begin{equation}\label{25}
\gr \left(  n_{\nu_{l}} \; - \; n_{\ov{\nu}_{l}} \right) \; < \; 0 ~~~~.
\end{equation}
Summing up, we found that neutrino ponderomotive force is negative (i.e.
attractive towards the centre ($\propto - \bvec{r}$), assuming spherical
symmetry) on neutrons and positrons, while is positive (i.e. repulsive from
the centre) for protons and electrons.

\vspace{1cm}

Such a ponderomotive force exerted by neutrinos can play, in principle,  an
important role in Supernova dynamics. The major effect would be related to
the revival of the shock by means of the energy deposited in the background
material behind the shock itself. In the standard delayed mechanism, if
$\sigma$ is the averaged cross section for neutrino-electron and
neutrino-nucleon scattering,
\bea
\sigma(\nu e \rt \nu e) & \simeq & k \; G_F^2 \, E_\nu \, T \label{26} \\
\sigma(\nu_e n \rt e^- p) & \simeq & \omega_1 \; G_F^2 \, E_\nu^2
\label{27} \\
\sigma(\ov{\nu}_e p \rt e^+ n) & \simeq & \omega_2 \; G_F^2 \, E_\nu^2
\label{28}
\eea
(where $k$ is a given constant depending on the particular channel, while
$\omega_1$, $\omega_2$ are (smooth) functions of neutrino energy $E_\nu$;
see \cite{Tubbs} for more details), the energy gain of a given particle in
the background at a distance $R$ is \cite{Bethe}
\begin{equation}\label{29}
  \frac{dE}{dt} \; = \; \frac{{\cal L} \, \sigma}{4 \pi R^2}
\end{equation}
${\cal L}$ being the luminosity in neutrinos or antineutrinos. This
relation is not, in general, complete since the material absorbing the
neutrinos can also emit neutrinos spontaneously; the total energy change is
then given by (\ref{29}) times a correction factor estimated as
\cite{Bethe}
\begin{equation}\label{30}
  1 \; - \; \left( \frac{2 R}{R_\nu} \right)^2 \, \left( \frac{T}{T_\nu}
  \right)^6
\end{equation}
where $R_\nu$. $T_\nu$ are the radius and temperature of the neutrinosphere
(while $T$ is the material temperature). For typical numbers,
\bea
 {\cal L} & \approx & 5 {\times} 10^{52} \, ergs/s \nonumber \\
 < E_\nu^2 > & \approx & 6 T_\nu^2 \; \approx \; 100 \, MeV \label{31} \\
 R & \approx & 150 \, Km
\eea
and assuming that all nucleons are free behind the shock, the absorption
average energy gain per nucleon is (from (\ref{29}))
\begin{equation}\label{32}
  \frac{dE}{dt} \; \simeq \; 50 \, Mev/s
\end{equation}
while for neutrino-electron scattering the rate is about 1/2 of this, due
to the different cross section. \\ However, there are also collective
interactions of neutrinos on the background plasma described by the
ponderomotive force in (\ref{9})-(\ref{11}). The energy change induced on
$e^{\pm},p,n$ in the material by this force is obtained from
\begin{equation}\label{33}
  \frac{dE_\alpha}{dt} \; = \; \bvec{{\cal F}}_\nu^\alpha \, {\cdot} \,
  \bvec{v}_\alpha
\end{equation}
where $\bvec{v}_\alpha$ is the velocity in the medium of the considered
particle. By comparison of (\ref{33}) and (\ref{29}) we immediately see
that, as expected \cite{b2} \footnote{In \cite{b2} there is an
overestimation of the momentum transfer from the ponderomotive force to the
background, leading to exceedingly large values for the neutrino force.},
collective effects are in general more relevant that incoherent ones since,
while the rate in (\ref{29}) is proportional to $G_F^2$, the expression in
(\ref{33}) is of order $G_F$ and then the mechanism more efficient.
However, for the present case, there are also two important suppression
factors to be taken into account, namely the dependence of $\bvec{{\cal
F}}_\nu$ on the difference between neutrinos and antineutrinos number
densities (which would be almost equal for $\nu, \ov{\nu}$ produced
thermally by $e^+ e^-
\rt \nu \ov{\nu}$) and the dependence od $dE/dt$ on the average angle between
the ponderomotive force and the velocity of the background particle. Let us
first calculate the maximum energy gain implied by (\ref{33}); to this end
we have first to estimate the net neutrino number in a Supernova. We adopt
a very simple model in which spherical symmetry is assumed: neutrinos are
emitted, with a Fermi-Dirac distribution of temperature $T_\nu$, from the
neutrinosphere (practically coincident with the core surface) located at
$R_\nu$ and escape freely outward from that. No neutrino source is
considered outside the neutrinosphere, so that the neutrino flux decrease
with distance only for a dilution spatial factor,
\begin{equation}\label{34}
  n_{\nu_l} \; - \; n_{\ov{\nu}_l} \; = \; \left(
  n_{\nu_l} \; - \; n_{\ov{\nu}_l} \right)_0 \, \left(
  \frac{R_\nu}{r} \right)^2
\end{equation}
(for $r > R_\nu$) where $\left( n_{\nu_l} \; - \; n_{\ov{\nu}_l} \right)_0$
is the neutrino net number density at neutrinosphere, which we now pass to
evaluate. Let us first concern on non electron neutrinos, that are produced
entirely by thermal processes. The difference between neutrino and
antineutrino distribution arises in this case only from the (slightly)
different neutrino-nucleon elastic scattering cross section, and vanishes
in the infinite nucleon mass approximation. The net number is then
proportional to $T/M_N$ ($M_N$ is the nucleon mass) and is calculated in
\cite{charlie}; at the neutrinosphere we have
\begin{equation}\label{35}
  \left( n_{\nu_x} \; - \; n_{\ov{\nu}_x} \right)_0 \; \simeq \;
  \frac{\pi^2}{36} \, \delta \, \left( \frac{T_{\nu_x}}{M_N} \right)
  \, T_{\nu_x}^3
\end{equation}
($x = \mu, \tau$) where $\delta$ is a parameter depending on nucleon weak
couplings, equal to 3.32 for neutrons and 2.71 for protons. \\ The
difference in $\ne$ and $\ov{\nu}_e$ distributions induced by the different
neutral current scattering is the same as in (\ref{35}); however, for
electron neutrinos and antineutrinos there is a more important contribution
coming from charged current (elastic and inelastic) scattering. This
induces a (not large) effective degeneracy parameter $\eta_\nu = \mu_\nu /
T_\nu$, and the net electron neutrino number is approximately proportional
to this parameter:
\begin{equation}\label{36}
  \left(  n_{\nu_e} \; - \; n_{\ov{\nu}_e} \right)_0 \; \simeq \;
  \frac{1}{6} \, \eta_\nu \, T_\nu^3 ~~~~.
\end{equation}
Following \cite{Bethe}, $\eta_\nu$ can be estimated from  the fact that in
a Supernova the ratio of the (average) energies in $\ne$ and $\ne +
\ov{\nu}_e$, due to the deleptonization electron neutrinos from the core,
is approximately equal to 5/8. From this we have
\begin{equation}\label{37}
  \eta_\nu \; \simeq \; 0.29
\end{equation}
(but this result depends somewhat on $T_\nu$ and on the average neutrino
chemical potential in the core).\\ From the above arguments we are now able
to evaluate quantitatively the ponderomotive force exerted by neutrinos; we
get \footnote{For illustrative purposes, we take the value 3.32 for
$\delta$ and consider $e$-neutrinosphere and $x$-neutrinosphere as located
at about the same radius $R_\nu$, although with different temperature. This
last approximation is justified from the fact that we will concern with
effects induced by the ponderomotive force taking place not very near the
neutrinospheres. Obviously, both these assumptions have to be relaxed in a
detailed Supernova numerical simulation.}
\bea
\fem & = & - \, \fep \; \simeq \nonumber \\
& \simeq & 2.94 {\times} 10^{-6} \left\{
\left( \frac{\eta_{\nu_e}}{0.29} \right)  \, \left( \frac{T_{\nu_e}}{4 \, MeV}
\right)^3 \; - \; 3.2 {\times} 10^{-2} \, \left( \frac{\delta}{3.32} \right) \,
\left( \frac{T_{\nu_x}}{6 \, MeV} \right)^4 \right\} \, {\cdot} \nonumber \\
& {\cdot} & \left( \frac{10 \, Km}{R_\nu} \right) \, \left( \frac{R_\nu}{r}
\right)^3 \; \hat{\bvec{r}} \;\;\; \frac{MeV}{c} \, s^{-1} \label{38} \\
\fp & \simeq &  1.2 {\times} 10^{-7} \left\{
\left( \frac{\eta_{\nu_e}}{0.29} \right)  \, \left( \frac{T_{\nu_e}}{4 \, MeV}
\right)^3 \; + \; 0.79 \, \left( \frac{\delta}{3.32} \right) \,
\left( \frac{T_{\nu_x}}{6 \, MeV} \right)^4 \right\} \, {\cdot} \nonumber \\
& {\cdot} & \left( \frac{10 \, Km}{R_\nu} \right) \, \left( \frac{R_\nu}{r}
\right)^3 \; \hat{\bvec{r}} \;\;\; \frac{MeV}{c} \, s^{-1} \label{39} \\
\fn & \simeq &  - \, 1.5 {\times} 10^{-6} \left\{
\left( \frac{\eta_{\nu_e}}{0.29} \right)  \, \left( \frac{T_{\nu_e}}{4 \, MeV}
\right)^3 \; + \; 0.79 \, \left( \frac{\delta}{3.32} \right) \,
\left( \frac{T_{\nu_x}}{6 \, MeV} \right)^4 \right\} \, {\cdot} \nonumber \\
& {\cdot} & \left( \frac{10 \, Km}{R_\nu} \right) \, \left( \frac{R_\nu}{r}
\right)^3 \; \hat{\bvec{r}} \;\;\; \frac{MeV}{c} \, s^{-1} \label{40}
\eea
Note that, for a given distance, the force acting on electrons is about a
factor 13 greater than the one acting on protons, while electrons and
neutrons experience about the same force. Observe also that while the
contribution of $\nm, \nt$ is subdominant with respect to that of $\ne$ in
$\fem$, each flavour of neutrinos contribute with about the same strength
in $\fp, \fn$. \\ We can now estimate the maximum energy gain of the
Supernova material behind the shock front due to the considered collective
effects by assuming that electrons are relativistic while the mean nucleon
velocity is given by $\sim \sqrt{3 T / M_N} \approx 0.06$. For $R = 150 \,
Km$ we clearly find that the electron energy gain is about a factor $3 {\times}
10^{-11}$ less than that of the standard results, and the electron heating
is the major effect for collective interactions, contrary to the standard
picture in which the nucleon heating (by inelastic scattering) is the most
remarkable one. Moreover, in the proposed scenario, another suppression
factor comes from the dependence on the angle between electron (or nucleon)
motion and the ponderomotive force (see Eq. (\ref{33})). In fact, the
previous estimate of the heating rate assumes that nearly all the electrons
(or nucleons) are polarized; then, although the magnetic field contribution
to $\bvec{{\cal F}}_\nu$ is unexisting (as established here), it would be
nevertheless important for reaching the maximum effect \footnote{In
\cite{magn2} the mean electron polarization in a Supernova for the Wilson
\& Mayle model has been calculated. The authors find that for $r
\simeq 100 \div 200 \, Km$ it can be very high (near the unit value)
allowing the magnetic field in this region to be as high as $\sim 10^{16}
\, Gauss$.}, which is, however, completely subdominant with respect to the
standard heating mechanism, the heating rate being some $eV \, s^{-1}$ for
$r = R_\nu$ and decreasing as $r^{-3}$. \\ Nonetheless, the action of the
ponderomotive force due to neutrinos is not, in general, restricted to heat
the material behind the shock (at least in a direct way). It could give a
non negligible contribution to the theory of convection and instabilities
during a Supernova explosion. In fact we remind, from
(\ref{38})-(\ref{40}), that while $\fem$ and $\fp$ are positive, $\fep$ and
$\fn$ are negative. This means that $e^-$ and $p$ are pushed in the
outgoing radial direction, while $e^+$ and $n$ in the ingoing one, and this
could be a relevant source of convection, as pointed out in \cite{pond},
especially near the radiation bubble, where Rayleigh-Taylor instabilities
already act.

\vspace{1cm}

In conclusion, we have studied collective interactions of a beam of
neutrinos/antineutrinos traversing a dense plasma of $e^{\pm}, p , n$ and
applied the results to the case of a Supernova. We have found that the
ponderomotive force exerted by neutrinos on the material behind the shock
cannot substantially heat that and gives a negligible contribution to the
revival of the shock for a successful Supernova explosion. In particular,
the magnetic field present in the considered plasma does not contribute to
the expression of the ponderomotive force, although it is relevant for the
polarization effects induced by it in the medium. \\ The collective
interactions studied here can, nevertheless, play a relevant role in dense
stellar plasmas \cite{land} where they can provide an additional plasma
cooling process through neutrino Landau damping of electron plasma waves,
thus influencing the evolution of massive stars.


\vspace{1truecm}

\noindent
{\Large \bf Acknowledgements}\\
\noindent
The author is deeply indebted with Dr. S. J. Hardy and Dr. L. O. Silva for
valuable discussions, and with Prof. V. S. Berezinsky for his remarks.


\begin{thebibliography}{99}

\bibitem{Bethe}
H.A. Bethe, {\it Rev. Mod. Phys.} {\bf 62} (1990) 801.

\bibitem{Burrows}
A. Burrows, J. Hayes and B.A. Fryxell, {\it Ap. J.} {\bf 450} (1995) 830;\\
A. Burrows, in  {\it Proc. 18th Texas Symposium on Relativistic
Astrophysics}, eds. A. Olinto, J. Frieman and D. Schramm (World Scientific,
1998);\\ A.Burrows, {\it Nucl. Phys.} {\bf A 606} (1996) 151.

\bibitem{Bruenn}
S.W. Bruenn, {\it Astrophys. Space Sci.} {\bf 143} (1988) 15.

\bibitem{Wilson}
J.R. Wilson, in {\it Numerical Astrophysics}, edited by J.M. Centrella,
J.M. LeBlanc and R.L. Bowers (Jones \& Bartlett, Boston, 1985) p. 422;\\
H.A. Bethe and J.R. Wilson, {\it Ap. J.} {\bf 295} (1985) 14.

\bibitem{Cooperstein}
J. Cooperstein, H.A. Bethe and G.E. Brown, {\it Nucl. Phys.} {\bf A 429}
(1984) 527.

\bibitem{Bingham}
R. Bingham, J.M. Dawson, J.J. Su and H.A. Bethe, {\it Phys. Lett.} {\bf A
193} (1994) 279;\\ S.J. Hardy and D.B. Melrose, {\it Phys. Rev.} {\bf D 54}
(1996) 6491; \\ L.O. Silva, R. Bingham, J.M. Dawson, P.K. Shukla, N.L.
Tsintsadze and J.T. Mendo\c{c}a, preprint physics/9807050.

\bibitem{b2}
R. Bingham, H.A. Bethe, J.M. Dawson, P.K. Shukla and J.J. Su, {\it Phys.
Lett.} {\bf A 220} (1996) 107.

\bibitem{pond}
L.O. Silva, R. Bingham, J.M. Dawson and W.B. Mori, preprint
physics/9807049.

\bibitem{BB}
G. Raffelt, {\it Stars as Laboratories for Fundamental Physics} (The
University of Chicago Press, 1996);\\ C. Thomson and R.C. Duncan, {\it Ap.
J.} {\bf 408} (1993) 194.

\bibitem{NR}
D. N\"{o}tzold and G. Raffelt, {\it Nucl. Phys.} {\bf B 307} (1988) 924.

\bibitem{magn}
S. Esposito and G. Capone, {\it Z. Phys.} {\bf C 70} (1996) 55;\\ P.
Elmfors, D. Grasso and G. Raffelt, {\it Nucl. Phys.} {\bf B 479} (1996)
3;\\ J.C. D'Olivo and J. Nieves, {\it Phys. Lett.} {\bf B 383} (1996) 87.

\bibitem{magn2}
H. Nunokawa, V.B. Semikoz, A.Yu. Smirnov and J.W.F. Valle, {\it Nucl.
Phys.} {\bf B 501} (1997) 17.

\bibitem{charlie}
C.J. Horowitz and Gang Li, preprint hep-ph/9809492.

\bibitem{Tubbs}
D.L. Tubbs and D.N. Schramm, {\it Ap. J.} {\bf 201} (1975) 467.

\bibitem{land}
L.O. Silva, private communications.

\end{thebibliography}
\end{document}